# Guided Mid-IR and Near-IR Light within a Hybrid Hyperbolic-Material/Silicon Waveguide Heterostructure

M. He[1], S. I. Halimi[2], T. G. Folland[1, 6], S. S. Sunku[3,5], S. Liu[4], J. H. Edgar[4], D. N. Basov[3], S. M. Weiss[2], J. D. Caldwell[1,2,*]

1, Department of Mechanical Engineering, Vanderbilt University, Nashville, TN 37212, USA. 2, Department of Electrical Engineering and Computer Science, Vanderbilt University, Nashville, TN 37212, USA. 3, Department of Physics, Columbia University, New York NY 10027, USA. 4, Tim Taylor Department of Chemical Engineering, Kansas State University, Manhattan, KS 66506, USA. 5, Department of Applied Physics and Applied Mathematics, Columbia University, New York NY 10027, USA. 6, Department of Physics and Astronomy, The University of Iowa, Iowa City, Iowa, 52242

**Abstract**. Silicon waveguides have enabled large-scale manipulation and processing of near-infrared optical signals on chip. Yet, expanding the bandwidth of guided waves to other frequencies would further increase the functionality of silicon as a photonics platform. Frequency multiplexing by integrating additional architectures is one approach to the problem, but this is challenging to design and integrate within the existing form factor due to scaling with the free-space wavelength. Here, we demonstrate that a hexagonal boron nitride (hBN)/silicon hybrid waveguide can enable dual-band operation at both mid-infrared (6.5-7.0 µm) and telecom (1.55 µm) frequencies, respectively. Our device is realized via lithography-free transfer of hBN onto a silicon waveguide, maintaining near-infrared operation, while mid-infrared waveguiding of the hyperbolic phonon polaritons (HPhPs) in hBN is induced by the index contrast between the silicon waveguide and the surrounding air, thereby eliminating the need for deleterious etching of the hBN. We verify the behavior of HPhP waveguiding in both straight and curved trajectories, and validate their propagation characteristics within an analytical waveguide theoretical framework. This approach exemplifies a generalizable approach based on integrating hyperbolic media with silicon photonics for realizing frequency multiplexing in on-chip photonic systems.

**Introduction:** Creating on-chip optical technologies for high-speed optical processing[1,2], optical communications[1,3,4] and chemical spectroscopy[5] has been a focus for optics research in recent decades. In the near-infrared (near-IR), integrated silicon photonics have arguably become the most important sub-field after decades of development[1,2,6]. One of the essential components of silicon photonics is the waveguide, which is to transmit and route optical signals. With demand for continuous increases in operational bandwidth, there is a desire to expand the operating frequency regime as multiplexing near-IR and mid-infrared (mid-IR) signals could provide unique applications for both signal processing and chemical sensing[7]. However, this is challenging with silicon, since accommodating longer wavelength modes requires expanding the size of the waveguide, which in turn would cause severe modal dispersion in the near-IR. Thus, an approach based on confining longer wavelengths of light to deeply sub-diffractional length scales in an architecture where modal wavelengths commensurate with those of the waveguide modes in the near-IR silicon components would be highly beneficial.

Hyperbolic polaritons (HPs) [8-11] offer the ability to compress long-wavelength free-space light to deeply sub-diffractional volumes and hence overcome the length-scale mismatch with guided modes that propagate in silicon waveguides designed for the near-IR. Such HPs are supported within highly anisotropic media where the permittivity along orthogonal axes are opposite in sign[12]. While originally probed within artificial metamaterials, the discovery of hyperbolicity in hexagonal boron nitride (hBN) demonstrates this property occurs in an expanding list of natural materials[10,13,14]. In hBN, these HPs result from the coupling of light to the bound ionic charge on a polar lattice, giving rise to hyperbolic phonon polariton modes (HPhPs)[15]. Guiding of HPhPs has been studied previously in two contexts using hBN, using patterned strips[16-18] and through induction of in-plane refraction of HPhPs realized via local variations in the refractive index of the environment (e.g., substrate)[19,20]. In the context of the former, hBN strips were constructed through nanofabrication and physical etching processes. However, such etching introduces unavoidable material damage[8,16,21-23], increasing the loss, and thus limiting the propagation lengths of the modes. Further, within the frequency multiplexing approach outlined in this work, the presence of the silicon waveguide complicates the processing for subsequent patterning of the hBN architecture. The latter approach employs inducing refractive effects in plane[19,20], where patterning of the substrate or domains in phase-change materials can be used to control HPhP propagation for novel devices that were first theoretically proposed[19] and then experimentally reduced to practice[20]. While this approach involves no patterning and hence no degradation of the hyperbolic media, the use of phase change materials results in detrimental substrate absorption, which precludes their use as a waveguide medium in the near-IR. Thus, to realize frequency multiplexing, integrating chemical sensing platforms, and/or signal processing approaches, demonstrating the applicability of waveguide physics to the deeply sub-wavelength HPhPs requires a different approach.

Here we address this challenge by realizing guided mid-IR and near-IR light within a hybrid hyperbolic-material/silicon waveguide heterostructure (Fig. 1a). By exploiting the in-plane refraction that occurs when the volume-confined HPhPs propagate over a substrate domain between high and low refractive indices, we demonstrate that the mid-IR HPhPs can be guided in a planar slab of hBN without any fabrication to the slab itself. This is realized as the modes are induced to follow the underlying patterned silicon waveguide, which in turn can simultaneously support near-IR waveguide modes. Using scattering-type scanning near-field optical microscopy (s-SNOM) in the mid-IR, we experimentally observe these HPhP waveguide (HPhP-WG) modes, with both fundamental and higher-order modes reported. We corroborate this by showing the strong quantitative agreements of their spectral dispersion and modal profiles with analytical waveguide theory. Additionally, we validate the waveguide nature by demonstrating that the HPhP-WG modes propagate along sub-wavelength radii curved trajectories within ring waveguides as well. Finally, we experimentally show that the presence of the hBN slab results in negligible influence on waveguiding of the near-IR light, and thus, that the prototype hybrid waveguide can operate in both the near-IR and mid-IR simultaneously. Such heterogeneous, yet simplified, integration offers a generalizable approach for multiplexing dramatically different free-space wavelength light within a compact and on-chip footprint, offering a new toolset for nanophotonic design and fabrication that avoids damage to the hyperbolic materials.

**Fabrication and near-field measurements**

Because the hybrid system is based on the relatively mature silicon photonics platform, maintaining the architecture of existing silicon waveguides in the near-IR can save significant redesign. Thus, the samples employed here use pre-patterned silicon waveguides fabricated on silicon-on-insulator (SOI) wafers with a 220-nm thick device layer, a common choice for silicon photonics. For mid-IR operation, $^{10}$B enriched hBN (~99% enriched[24,25]) flakes were exfoliated and transferred onto the pre-patterned SOI silicon waveguides using low contamination transfer techniques[26]. The hBN crystals were grown with a boron source that was nearly 100% $^{10}$B isotope, as previously described[27].

Using s-SNOM with mid-IR excitation from a line-tunable quantum cascade laser (QCL), spatial mapping of the evanescent optical fields on the structure can be directly monitored at the defined incident laser frequency. Consistent with prior works[28,29], fringes resulting from interference effects between the launched and propagating HPhPs enables direct probing of the compressed HPhP wavelength ($\lambda_{HPhP}$), which propagates primarily within the volume of the hBN slab (Fig. 1a). In s-SNOM images, HPhPs can be observed in two ways: first, polaritons launched by light scattered from the s-SNOM tip propagate radially away from the tip and reflect back from sample boundaries (e.g., a flake edge) creating interference fringes with spacing $\lambda_{HPhP}/2$ [9,28,29], which are scattered back to free space by the tip and detected. Alternatively, polaritons can be directly launched from the edge of patterned silicon structures under hBN or the edge of the hBN flake and then propagate across the surface and interfere with the incident field at the tip, producing fringes with spacing $\lambda_{HPhP}$ [19,24,30].

**Hyperbolic phonon polariton waveguide formalism**

To form a waveguide mode, two criteria must be satisfied[31]: (i) total internal reflection and (ii) constructive interference of the optical mode within the waveguide. In our geometry, we have exfoliated and transferred hBN flakes onto pre-patterned silicon waveguides, with the surrounding region where silicon was etched away resulting in freely suspended hBN (Fig. 1a). The refractive index difference between silicon and air thus dictates the $\lambda_{HPhP}$ supported within hBN, with a two- to six-times difference in wavevector (or equivalently, effective refractive index) realized within the upper Reststrahlen band of hBN[9,15,32], as shown in Supplementary Information (SI), section 1. Such contrast in the refractive index is sufficient to induce total internal reflection of the HPhPs, with the reflections occurring inside the hBN at the boundaries between silicon and air. In order to satisfy criterion (ii), we need to use the wave equation and boundary conditions at the interface. The out-of-plane electric field ($E_z$) of propagating HPhPs share the same form of wave equation as in non-polariton material systems[9]:

$$E_z = E_0 \times e^{i(\mathbf{k}\mathbf{r}-\omega t)} \qquad (1)$$

where $E_0$ is the amplitude, $\mathbf{k}$ is the in-plane wavevector, $\mathbf{r}$ is the position vector, $\boldsymbol{\omega}$ is the frequency of the incident light and $t$ is time. Prior research has shown that the in-plane propagating hyperbolic waves follow Snell's law[19,20], which implies the continuity of the tangential component of the in-plane wavevector. Because the propagating HPhPs share the same form of the wave equation and presumably the same boundary conditions with dielectric materials, we anticipate that within our heterostructure design, a HPhP waveguide (HPhP-WG) in the x-y plane will result. Since the electric field is in the z-direction, and it is propagating in the y-direction, we can treat the system as a TE dielectric slab waveguide. In detail, the HPhP-WG modes ($\beta$) are the eigenvalues of the characteristic equation for the TE modes of a slab waveguide:

$$\tan(\kappa \times d) = \frac{2\kappa\gamma}{\kappa^2 - \gamma^2} \qquad (2)$$

where $\kappa = \sqrt{k_f^2 - \beta^2}$ and $\gamma = \sqrt{\beta^2 - k_s^2}$, and $d$ is defined as the width of the underlying silicon waveguide, and $k_f$ and $k_s$ are complex-valued wavevectors of HPhPs supported over the silicon and suspended regions, respectively (details of the HPhP-WG derivation are included in SI, section 3).

Thus, the mid-IR HPhPs confined within the thickness of hBN (the z-axis) are guided within the waveguide region (the x-axis) and propagating in the y-axis. This does not influence the silicon waveguide underlying hBN, and the near-IR light remains confined within the silicon cross-section (x-z plane) and propagating in the y-axis, as shown in Fig. 1a.

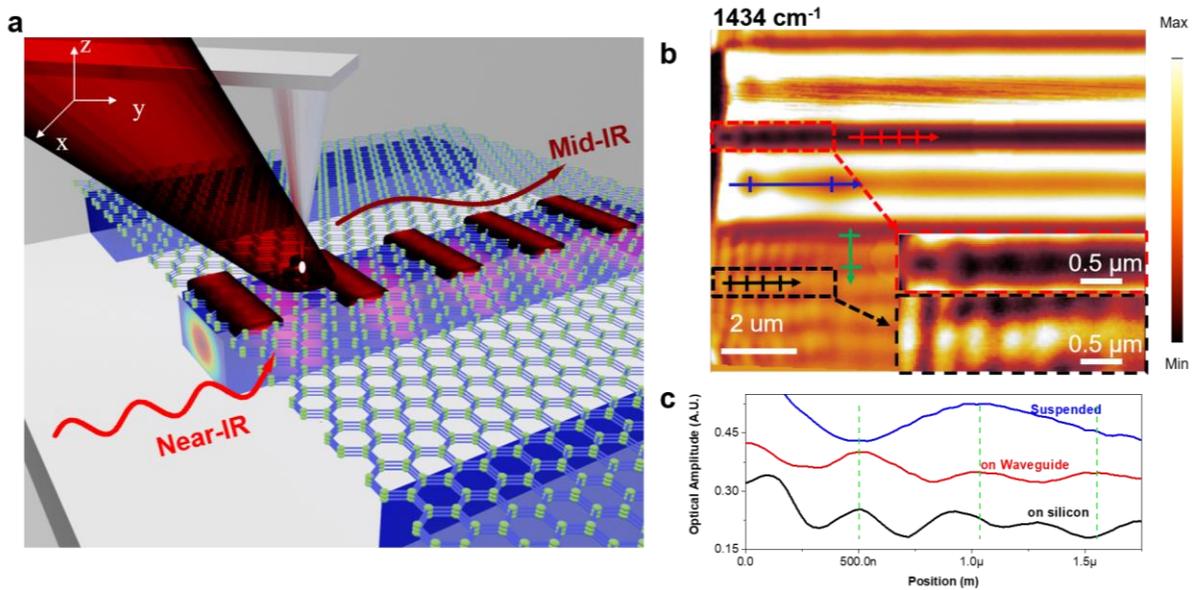

**Fig 1: Waveguiding of light in the mid-IR and near-IR within the same device. a,** A device and experimental schematic. The silicon waveguide is formed by etching away part of the silicon, and hBN is placed on top of it, resulting in hBN on waveguide surrounded by free-standing suspended hBN between silicon waveguide and unremoved silicon (blue boxes). HPhP-WG modes (dark red waves) are imaged by s-SNOM, while the transmission of near-IR light (pink modes in silicon) is not influenced, and the modal wavelengths of the two modes are nearly the same. **b,** s-SNOM amplitude image of hBN above a silicon waveguide at 1434 cm$^{-1}$, which shows the difference between HPhPs on silicon region and HPhP-WG. The red arrow is the HPhP-WG mode, the blue arrow is tip-launched HPhPs on suspended hBN, the black arrow is tip launched HPhPs on hBN above silicon, and the green arrow is edge launched HPhPs on hBN above silicon. The inset s-SNOM images are zoomed in images for HPhP-WG and hBN on silicon respectively, and they are at the same geometrical scale bar. **c,** Line scans of three different HPhPs: HPhP-WG, HPhPs on silicon region, and HPhPs on suspended region.

**Comparison between HPhP-WG and HPhPs**

To experimentally confirm the existence of HPhP-WG modes and compare the difference between HPhP-WG and normal HPhPs, we transferred a h$^{10}$BN flake onto a 0.7-μm-wide silicon waveguide (see optical images in SI, Fig. S4a ) and employed pseudo-heterodyne s-SNOM measurements on this prototype device at 1434 cm$^{-1}$ (6.97 μm free-space wavelength) to measure the propagating HPhP-WG modes, as shown in Fig. 1b. In this image, we can identify several tip-launched HPhPs within the hBN over the unpatterned silicon (black arrow) and air (suspended region, blue arrows), as well as edge-launched modes over the unpatterned silicon (green arrow). Additionally, in the region over the silicon waveguide, there is a tip-launched HPhP-WG mode (red arrow). To certify this is a waveguide mode, the line scans of the tip-launched HPhPs over the (1) unpatterned silicon, (2) suspended regions and (3) the silicon waveguide were extracted and plotted in Fig. 1c. As predicted by waveguide theory, the wavevector in the waveguide region (notated as β) is smaller than the wavevector of the HPhPs of the 'core' waveguide material ($k_f$, HPhPs over silicon), yet larger than the surrounding 'cladding' material ($k_s$, HPhPs over the suspended region). This prediction is confirmed by the line scans in Fig. 1c, as from inspection of the zoomed regions of interest included as insets in Fig. 1b.

**Mode analysis of HPhP-WG**

While the above-referenced prototype is suitable to identify the difference between HPhP-WG and HPhPs dictated simply by the substrate properties, the tip-launched nature of the excitation of the HPhP-WG modes in this geometry leads to the observation of the halved wavelength[9,28,29], making measurements more challenging due to the higher resolution that such detailed analysis requires. To better characterize the HPhP-WG system, we fabricated new silicon waveguides with a silicon-air edge (see SEM in Fig. S5a and optical image in Fig. S4b). This edge could scatter the incident mid-IR light, thus providing the necessary momentum to stimulate edge-launched HPhPs[19,24,30] into the waveguide. This enables direct observation of the HPhP-WG modes as shown in Fig. 2 a-f. Here, near-field images were collected at a variety of mid-IR frequencies, thus enabling the HPhP-WG dispersion relation to be extracted and compared to analytical theory.

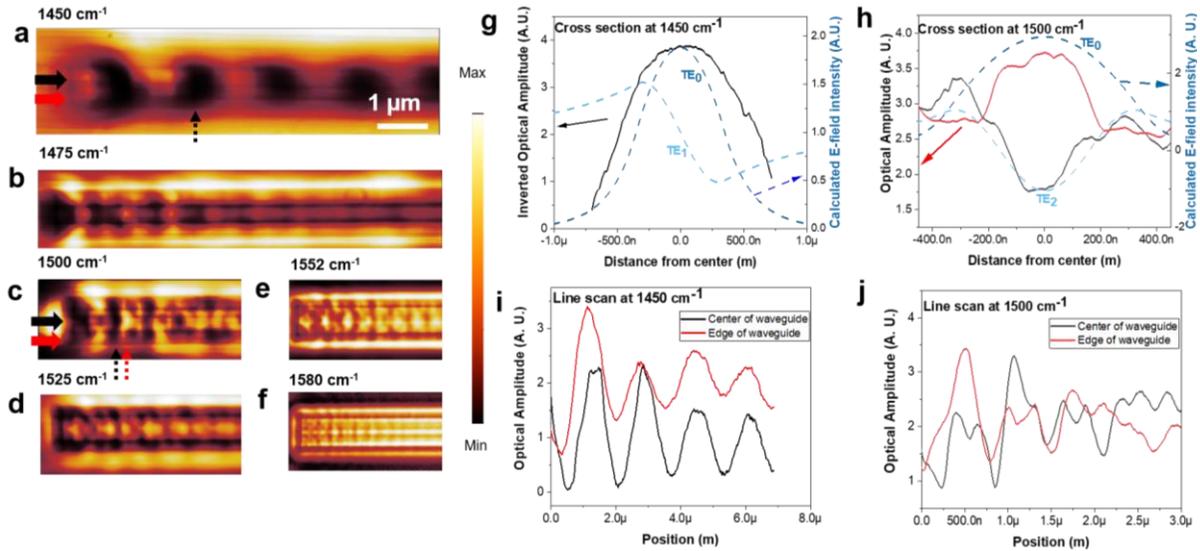

**Fig. 2: HPhP-WG mode analysis**. **a-f** Near-field optical amplitude images of hBN on a silicon waveguide at different frequencies. The geometry scale bars are the same and the near field intensity is plotted with non-linear scale implemented in Gwyddion software. **g, h,** Cross-sectional line scans at different lateral positions of waveguide at 1450 cm$^{-1}$ (1500 cm$^{-1}$). The line profiles are plotted in solid lines and dashed arrows in Fig. 2a, c indicate where they are taken, and calculated mode profiles are plotted in dashed lines. To counter the hyperlensing effect at 1450 cm$^{-1}$, a constant minus optical amplitude is plotted. **i, j,** Line scans along the waveguide at different vertical positions of waveguide at 1450 cm$^{-1}$ (1500 cm$^{-1}$). Black (red) solid arrows in Fig. 2a, c show where line scans in the center (on the edge) are taken.

Although the HPhP-WG modes are conceptually visualized, additional modal analysis is necessary for more comprehensive understanding. Accordingly, the cross-sectional line scans at 1450 cm$^{-1}$ and 1500 cm$^{-1}$ are extracted and plotted against the calculated modal profile, as shown in Fig. 2g and h. Because we are unavoidably imaging the underlying silicon waveguide through hBN via the hyperlensing effect (see more discussions in SI, section 6), the optical amplitude within the waveguide region is consistently lower than the outside region at 1450 cm$^{-1}$. To compensate for this effect, we plot the inverted optical amplitude (a constant minus optical amplitude) instead of raw optical amplitude, and the inverted line scan is indeed similar to the analytically calculated fundamental TE mode (Fig. 2g). At 1500 cm$^{-1}$, the $k_s$ and $k_f$ are both increased, and additional modes are allowed. We propose that all allowed modes should be excited and superimposed upon each other because HPhPs are launched into the waveguide via the silicon-air interface without modal selectivity. Indeed, there are two distinct cross-sectional scans: one that appears like the fundamental TE mode (TE$_0$), and another that exhibits a field profile similar to 2$^{nd}$-order TE mode (TE$_2$), which shows two peaks near the edge of the waveguide. Therefore, we assign these HPhP-WG modes as the TE$_0$ and TE$_2$, while the antisymmetric mode (TE$_1$) is not observed. Further, the difference between modes can be revealed by examining line scans along the waveguide propagation direction at different spatial positions[16]. The analytical model predicts that the fundamental TE$_0$ is highly confined within the center of the waveguide; in contrast, the TE$_2$ HPhP-WG mode exhibits high amplitudes near the waveguide edge (Fig. 2 g, h). At 1450 cm$^{-1}$, the experimental line scans extracted from these two locations exhibit field profiles that are nominally the same, with the exception that the line scan taken near the edge shows smaller signal variation, caused by the confinement of the TE$_0$ mode. Consistent with the cross-sectional analysis, we observe two different patterns at 1500 cm$^{-1}$. Within the center of the waveguide, the line scan is dominated by the TE$_0$ mode (high-frequency signal), while near the edge the TE$_2$ mode (low-frequency signal) is stronger. Thus, the s-SNOM line scans offer strong qualitative agreements with the trends anticipated from analytical modeling, providing strong support for our identification of the observed modes as TE$_0$ and TE$_2$. These field profiles were further confirmed by three-dimensional finite-element simulations (SI, section 7). Note that neither in the experimental data nor the three-dimensional simulations, do we observe the presence of antisymmetric modes. We believe that TE$_1$ and TE$_3$ modes (and other odd modes) are not excited in our geometry due to the antisymmetric (Fig. 2g) field profiles, which are incompatible with HPhP-WGs launched by the symmetric waveguide edge. In the three-dimensional simulations, a point dipole was used to stimulate the HPhPs, which exhibits similar symmetry as the edge, again precluding the generation of the antisymmetric modes. Although we do not experimentally detect those modes, we verify their existence from numerically calculated eigenmodes (SI, section 8), and do state that there is no inherent reason why they cannot be supported within other design geometries and/or with alternative excitation schemes.

**Dispersion of HPhP-WG modes**

While analyzing the s-SNOM images offers excellent qualitative agreement with both analytical calculations and three-dimensional simulations for the HPhP-WG mode profiles at several discrete frequencies, the wavevector analysis of HPhP-WG modes in the heterostructure system is required to validate the degree of quantitative agreement. Predictive capabilities of the HPhP-WG wavevector are imperative for developing more advanced applications, such as understanding out-of-plane coupling in chemical sensing concepts and for the design of more complicated waveguide geometries, such as one-dimensional photonic crystal structures[5,33]. With the same analytical model, the β of the allowed modes at each discrete frequency are calculated and plotted as dashed lines in Fig. 3. Similar to dielectric waveguides, at higher values of $k_s$ and $k_f$, multiple HPhP-WG modes are potentially allowed. More specifically, while the TE$_0$ mode is observed at all frequencies within the upper Reststrahlen band of hBN, the TE$_1$ is predicted at frequencies above 1445 cm$^{-1}$, and the onset of the TE$_2$ mode should occur at frequencies higher than 1495 cm$^{-1}$. By extracting the line profiles and fitting (details of fitting are included in SI section 9), we get the β of the HPhP-WG modes from the s-SNOM images (solid spheres) and directly compare these to the analytical calculations (dashed lines), while also including the results from the 3D-simulated field profiles (empty circles) in Fig. 3. Consistent with our prior line scan analysis, we find that the dispersion of the HPhP-WG modes is in excellent quantitative agreement with the analytically calculated TE$_0$ and TE$_2$ modes. While the HPhP-WG is propagating mid-IR light with a free-space wavelength on the order of 7μm, the HPhP-WG β is indeed comparable to that of the guided near-IR modes in the silicon waveguide at telecom frequencies ($\beta_{HPhP} \approx \beta_{Si} \approx 1 \times 10^5$ cm$^{-1}$). This design choice implies that the HPhP-WG in the mid-IR can be designed to share the propagation and modal characteristics with the silicon waveguide modes. Comparing with lossless silicon telecom counterpart, the fast dispersion and increased absorption loss of the hBN HPhP-WG modes certainly result in shorter propagation lengths, and the experimentally measured propagation length is limited to 10 μm (SI, section 10). However, its incorporation into the design demonstrates a possible means for frequency multiplexing within a single planar system that is not limited by scaling issues associated with simply increasing the number of silicon waveguides, while providing sufficient propagation[24] for inter- and intra-chip photonic connections. Furthermore, this offers a generalized approach towards expanding the operating frequency to any regime where hyperbolic media can be realized with reasonably low optical losses, independent of free-space wavelength. Finally, the use of mid-IR frequencies for the HPhP-WG operation also opens possibilities for chemical sensing applications for lab-on-a-chip approaches due to the spectral overlap with chemical vibrational fingerprint range through surface-enhanced infrared absorption (SEIRA)[34,35], strong-coupling[36,37] or refractive index sensing[38] modalities.

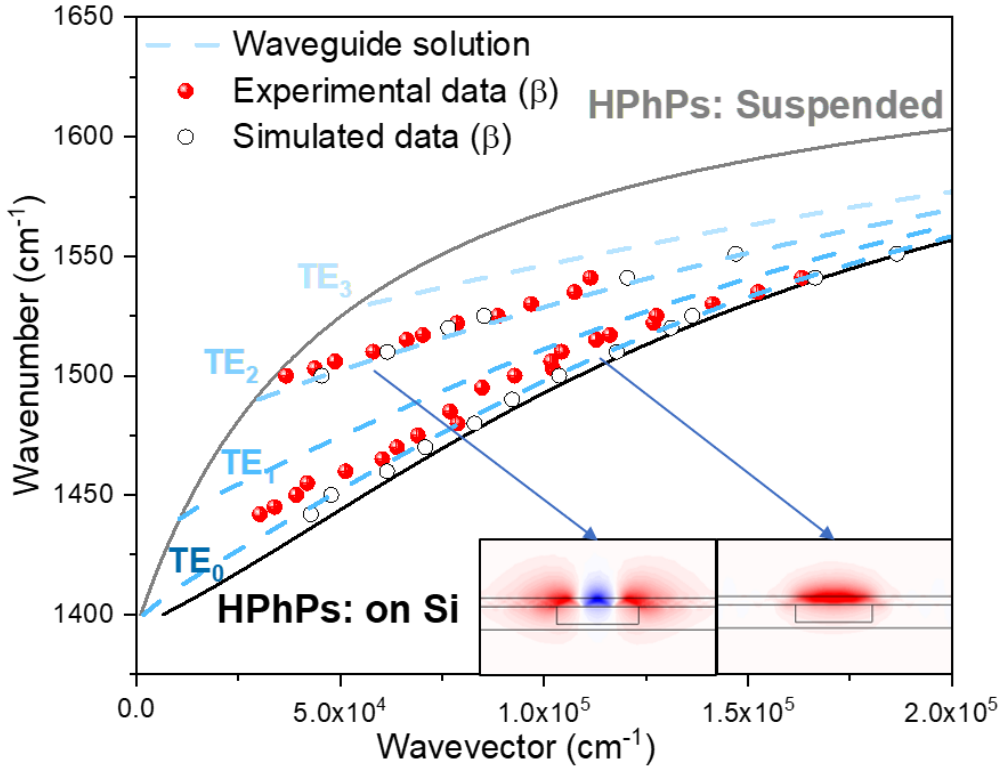

**Fig. 3: Dispersion curve of HPhP-WG modes**. The experimental $\beta$ are extracted from line scans of s-SNOM images and plotted as solid spheres, and the simulated $\beta$ are extracted in the same manner from three-dimensional simulations and plotted as empty circles. The dashed lines are analytically calculated mode solutions, and solid lines are calculated HPhPs of hBN on silicon region and suspended region respectively. Two numerically calculated mode profiles at 1508 cm$^{-1}$ are inserted to intuitively show the modes.

**HPhP-WG in non-linear trajectories**

In on-chip photonics, there is a requirement to be able to transmit information not only along linear trajectories, but also to propagate light around bends with varying degrees of curvature to route and process optical signals. To demonstrate that HPhP-WG modes can be guided around such curves, we also have investigated curved waveguides. The width of the curved waveguide was designed to be 0.6 µm, with a break in the ring structure included to serve as the launching point for the HPhPs. We fabricated structures featuring inner radii that are sub-wavelength in scale of 1, 2 and 3 µm in an effort to demonstrate propagation around different curvatures (Fig. 4d inset and additional SEM images in SI section 5). A single hBN flake was transferred[26] onto all three, so the data from these different structures could be compared directly (see the optical image in SI, Fig. S4c), without having to consider variations in the HPhP dispersion induced by variations in hBN thickness[9,15,39]. Again, we employ s-SNOM to characterize these waveguides and show that the HPhPs are indeed guided along the curved trajectories, with representative behavior for all three structures provided in Fig. 4a-c using an incident frequency of 1542 cm$^{-1}$. This guiding behavior is even maintained for ring radii and waveguide widths that are over six- and ten-times smaller, respectively, than the free-space wavelength ($\lambda_{FS}$ =6.49 µm), respectively. To ensure that these modes are HPhP-WG modes, we again compare the spectral dispersion of the analytically calculated β with the values extracted from these s-SNOM experiments (Fig. 4d). Although the β for the TE$_0$ modes extracted from different radii waveguides do vary slightly from each other, the agreement of all extracted wavevectors with the analytical calculations remains excellent. Thus, these modes are clearly consistent with HPhP-WG modes described above, illustrating the potential for routing mid-IR optical signals on-chip, even around bends. This demonstrates the potential to integrate this approach with designs relying on such structures, such as ring resonators and Mach-Zehnder interferometers.

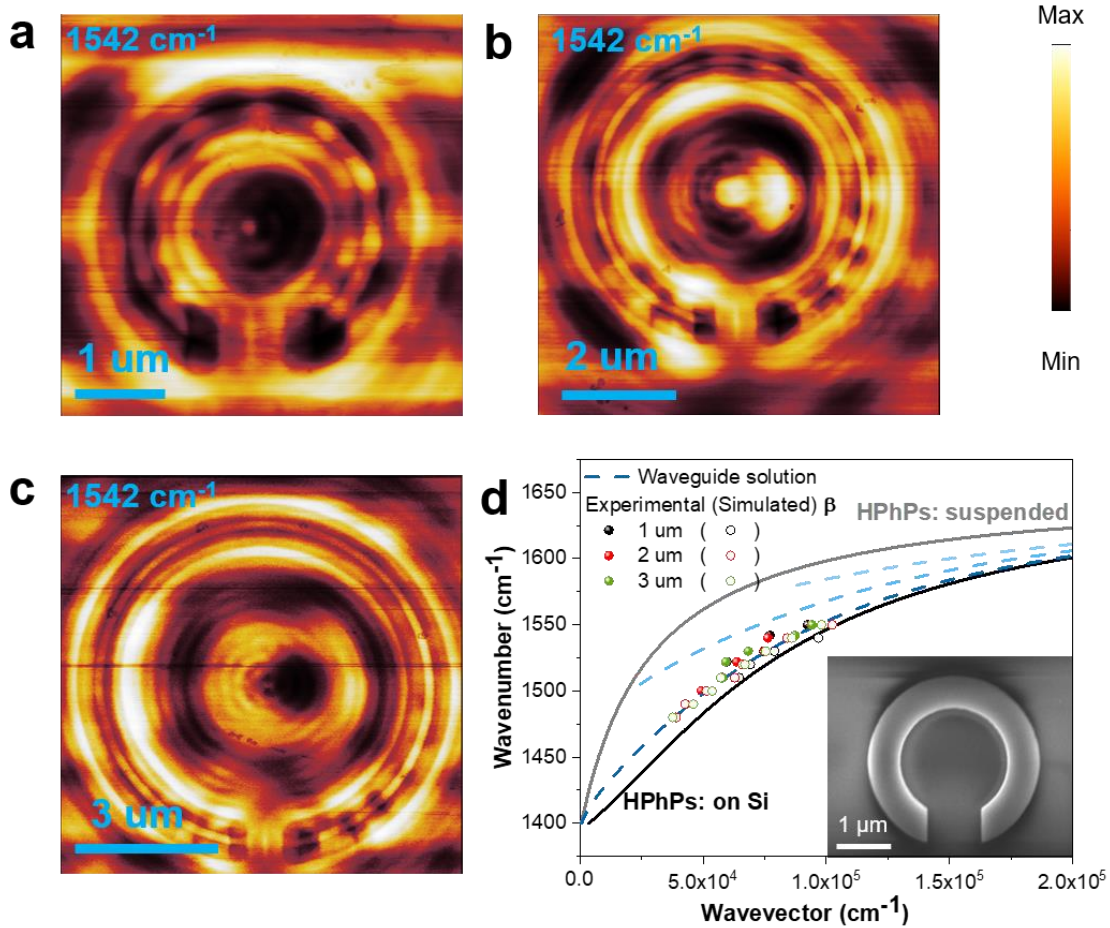

**Fig. 4**: **HPhP-WG modes in curved structures. a-c,** Near-field optical amplitude images of hBN on curved silicon waveguide structures at 1542 cm$^{-1}$. While the widths of the waveguide are the same (0.6 μm), the inner radii of the three structures are 1 μm, 2 μm, and 3 μm respectively. **d,** Dispersion curve of the HPhP-WG modes in curved structure with different radii. The experimentally extracted data are plotted in solid spheres, and the simulated $\beta$ are plotted in empty circles. The dashed lines are analytical solutions of slab waveguide model, and solid lines are calculated HPhPs of hBN on silicon and suspended hBN respectively. One representative SEM image of the designed curved waveguide is inserted.

**Performance of the hybrid system in the near-IR**

The potential for guiding long-wavelength mid-IR light through our hybrid HPhP-silicon waveguide platform opens significant opportunities for frequency multiplexing that could significantly expand operational bandwidth in on-chip silicon photonic approaches. However, while the HPhP-WG modes can indeed be described by analytical slab waveguide model and support HPhP wavelengths commensurate with the modal wavelengths of the silicon waveguide modes in the near-IR, this is only useful if the presence of the hyperbolic material is determined to be benign towards near-IR operation. As hBN has a low refractive index at near-IR frequencies, its effect on the silicon photonics platform at 1.55 μm should be minimal. To affirm this, we employ an electromagnetic eigenmode solver to determine the various allowed modes within a 0.7-μm wide silicon waveguide on SOI, with the primary guided modes presented in Fig. 5a, b (more allowed modes are included in SI section 11). These calculations are repeated for the waveguide covered with a 50-nm thick hBN, where it is determined that the presence of the hBN has minimal impact upon the profiles of the two guided modes. More specifically, the effective modal index is changed by 2.0% (5.0%) for the $0^{th}$ ($1^{st}$) mode, as shown in Fig. 5 c, d. To experimentally validate this, hBN flakes were transferred onto a silicon waveguide sample, and the transmission intensity was measured before and after the hBN was transferred (Fig. 5e). Although we see immeasurable changes from simulations (Fig. S12), the transmission of the silicon waveguide partly covered with hBN is reduced slightly. Because hBN is lossless in the near-IR, the only way that the existence of hBN influences the transmission is to induce reflection and scattering at the edge of hBN. To find out how severe the scattering is, we collected IR images when the light was coupled into the waveguide. The infrared camera image is overlaid with the edge of hBN extracted from the visible image to show the position of the hBN flake, as shown in the inset of Fig. 5e (raw data are included in SI, section 12). Even though we see significant scattering on the waveguide that could be caused by wear on the chip and surface roughness, the scattering where the light enters the hBN/waveguide region is negligible, which is consistent with simulations (Fig. S12). Thus, we attribute the reduction in transmission intensity to minor degradation of the chip during handling and a possible slight variation in coupling efficiency for the two measurements rather than the presence of hBN, and we believe that a pristine hBN layer on top of a silicon waveguide would have negligible influence in the near-IR.

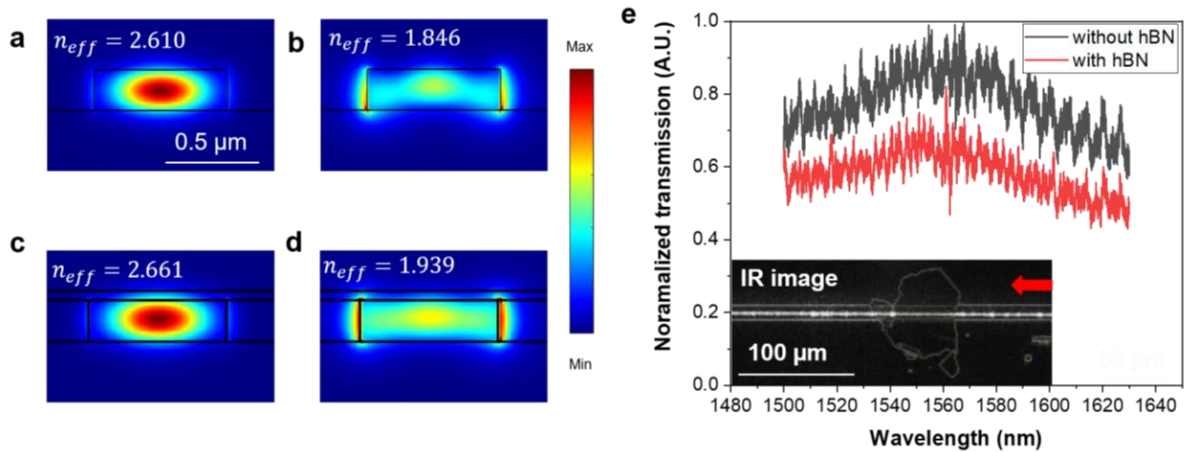

**Fig. 5: Performance of the hybrid system in the near-IR. a, b,** Mode profiles (intensity of electrical field) at 1550 nm for $0^{th}$- and $1^{st}$-order modes in a silicon waveguide without hBN. **c, d,** Equivalent mode profiles (intensity of electrical field) at 1550 nm for a silicon waveguide with hBN on top. All mode profiles here are at the same geometrical scale bar. **e,** Measured transmission spectrum before and after hBN was transferred onto the silicon waveguide. The left inset shows an infrared camera image of a segment of the waveguide partly covered by a hBN flake, with light coupled in from a tunable laser (red arrow indicates the input port), and the image is overlaid with the edge of the hBN extracted from a visible image to show the position of hBN.

**Conclusion**

In summary, we have demonstrated a novel approach for frequency multiplexing of near-IR and mid-IR signals through a hybrid, hyperbolic-silicon photonic waveguide platform. Within this geometry, we have illustrated that two disparate frequencies of light can be guided within the same structure with comparable wavevectors. Benefitting from the nature of HPhPs, which strongly confines light to deeply sub-diffractional dimensions, yet remains sensitive to the local refractive index, the utilization of the patterned substrates to create HPhP-WGs in hBN becomes possible. We demonstrate that both the fundamental and high order HPhP-WG modes can be supported within a 0.7-μm wide hBN/silicon heterostructure without the need for patterning of the hBN, despite a free-space wavelength that is approximately ten times larger than the waveguide width. Moreover, we show that these HPhP-WG modes can be routed on-chip, exhibiting propagation around bends in ring waveguide designs featuring radii up to six-fold smaller than the free-space wavelength, further validating the waveguide nature of the mode and illustrating its utility. Finally, these HPhP-WG modes in the mid-IR can in principle be supported simultaneously with the silicon waveguide mode in the near-IR, as the presence of the hBN exhibits negligible impact upon near-IR light propagation for pristine samples. This, therefore, uncovers a potential generalizable solution for expanding operational bandwidth with minimal cross-talk due to the large spectral mismatch between signals, but within a single nanoscale integrated platform. Moreover, this approach of exploiting the underlying patterned substrates instead of the hyperbolic material offers a generic way to manipulate HPhPs without deleterious impacts upon optical loss.

## Acknowledgments

Work by J.D.C., M.H. and T.G.F. was supported by the Office of Naval Research under grant number N00014-18-1-2107 and via Vanderbilt University through J.D.C's startup package. Exfoliation and transfer of hBN were performed within the Vanderbilt Institute of Nanoscale Science and Engineering (VINSE) cleanroom, while silicon waveguide fabrication was conducted at the Center for Nanophase Materials Science, which is a DOE Office of Science User Facility. S.M.W. and S.I.H. were supported in part by the National Science Foundation (ECCS1809937). D.M.B. and S.S. were supported by the Air Force Office of Scientific Research under grant FA9550-15-1-0478, and D.M.B. is the Vannevar Bush Faculty Fellow (ONR-VB: N00014-19-1-2630). J.H.E. was supported by Materials Engineering and Processing program of the National Science Foundation (number CMMI 1538127).

## Methods

**Device Fabrication.** All the silicon waveguides were fabricated on a silicon-on-insulator (SOI) wafer with a 220 nm silicon device layer and a 3 μm buried oxide layer. We patterned the structured with standard nanofabrication procedures, using electron beam lithography (JEOL 9300FS) and reactive ion etching (Oxford Plasmalab 100) with a $C_4F_8$/$SF_6$/Ar gas mixture to define the waveguide geometries. The aforementioned lithography steps were performed at the Center for Nanophase Materials Sciences at Oak Ridge National Laboratory.

**Numerical simulations.** Three-dimensional finite-element simulations in the mid-infrared were conducted in CST Studio Suite 2018 using the frequency domain solver with open boundary conditions. In these simulations, polariton modes are only launched by a current point dipole above hBN, and field profiles are extracted using frequency monitors. All results used thicknesses consistent with that measured in topographic maps of the samples. Numerically calculated eigenmodes of the HPWG-WG were conducted in COMSOL. Modes for the silicon waveguides were calculated in the near-infrared in Lumerical MODE Solutions using a finite-difference eigenmode solver. Dielectric functions were taken from ref. [24] for isotopically enriched hBN.

**Near-field measurements.** Near-field nano-imaging experiments were carried out in a commercial (www.neaspec.com) s-SNOM based around a tapping-mode atomic force microscope. A metal-coated Si-tip of apex radius R $\approx$ 20 nm that oscillates at a frequency of $\Omega \approx$ 280 kHz and tapping amplitude of about 100 nm is illuminated by monochromatic quantum cascade laser beam at a wavelength λ = 6.9 μm and at an angle 60° off normal to the sample surface. Scattered light launches HPhPs in the device and the tip then re-scatters light (described more completely in the main text) for detection in the far-field. Background signals are efficiently suppressed by demodulating the detector signal at the third harmonic of the tip oscillation frequency and employing pseudo-heterodyne interferometric detection.

**Near-IR measurements.** The near-IR propagation characteristics were measured using TE-polarized light (electric field polarized across the waveguide) from a tunable semiconductor laser source that was end-fire coupled into the waveguides through tapered fibers. The transmission intensity, captured by a fiber-coupled optical power meter, demonstrates some reduction in transmitted power after the transfer of an hBN flake atop the sample, likely due to handling-induced wear.

## Author contributions

M.H. and J.D.C. conceived the idea. M.H. performed theoretical analysis. M.H., T.G.F. and S.I.H. performed the device fabrication. M.H. and T.G.F. performed mid-IR simulations. M.H. and S.I.H. performed the near-IR simulations. S.I.H. performed the near-IR measurements. M.H., T.G.F. and S.S.S. performed the s-SNOM measurements. S.L. and J.H.E. grew the hBN crystals. All authors analyzed the data and wrote the paper together.

# Reference

1       Willner, A. E., Khaleghi, S., Chitgarha, M. R. & Yilmaz, O. F. All-optical signal processing. *Journal of Lightwave Technology* **32**, 660-680 (2013).
2       Reed, G. T., Mashanovich, G., Gardes, F. Y. & Thomson, D. Silicon optical modulators. *Nature photonics* **4**, 518 (2010).
3       Patterson, D., De Sousa, I. & Achard, L.-M. The future of packaging with silicon photonics. *Chip Scale Review* **21** (2017).
4       Thomas, S., Bastardo, K. Y. & Anselm, M. K. in *2017 Pan Pacific Microelectronics Symposium (Pan Pacific).*  1-8 (IEEE).
5       Estevez, M. C., Alvarez, M. & Lechuga, L. M. Integrated optical devices for lab‐on‐a‐chip biosensing applications. *Laser & Photonics Reviews* **6**, 463-487 (2012).
6       Cheben, P., Halir, R., Schmid, J. H., Atwater, H. A. & Smith, D. R. Subwavelength integrated photonics. *Nature* **560**, 565-572 (2018).
7       Zou, Y., Chakravarty, S., Chung, C.-J., Xu, X. & Chen, R. T. Mid-infrared silicon photonic waveguides and devices. *Photonics Research* **6**, 254-276 (2018).
8       Caldwell, J. D. *et al.* Sub-diffractional volume-confined polaritons in the natural hyperbolic material hexagonal boron nitride. *Nature communications* **5**, 5221 (2014).
9       Dai, S. *et al.* Tunable phonon polaritons in atomically thin van der Waals crystals of boron nitride. *Science* **343**, 1125-1129 (2014).
10      Sun, J., Litchinitser, N. M. & Zhou, J. Indefinite by nature: from ultraviolet to terahertz. *Acs Photonics* **1**, 293-303 (2014).
11      Korzeb, K., Gajc, M. & Pawlak, D. A. Compendium of natural hyperbolic materials. *Optics express* **23**, 25406-25424 (2015).
12      Poddubny, A., Iorsh, I., Belov, P. & Kivshar, Y. Hyperbolic metamaterials. *Nature photonics* **7**, 948 (2013).
13      Álvarez-Pérez, G. *et al.* Infrared permittivity of the biaxial van der Waals semiconductor α-MoO$_3$ from near-and far-field correlative studies. *arXiv preprint arXiv:1912.06267* (2019).
14      Taboada-Gutiérrez, J. *et al.* Broad spectral tuning of ultra-low-loss polaritons in a van der Waals crystal by intercalation. *Nature materials*, 1-5 (2020).
15      Caldwell, J. D. *et al.* Photonics with hexagonal boron nitride. *Nature Reviews Materials* **4**, 552-567 (2019).
16      Dolado, I. *et al.* Nanoscale Guiding of Infrared Light with Hyperbolic Volume and Surface Polaritons in van der Waals Material Ribbons. *Advanced Materials*, 1906530 (2020).
17      Babicheva, V. E., Shalaginov, M. Y., Ishii, S., Boltasseva, A. & Kildishev, A. V. Long-range plasmonic waveguides with hyperbolic cladding. *Optics express* **23**, 31109-31119 (2015).
18      Maissen, C., Lopez, I. D. & Hillenbrand, R. in *2017 42nd International Conference on Infrared, Millimeter, and Terahertz Waves (IRMMW-THz).*  1-2 (IEEE).
19      Folland, T. G. *et al.* Reconfigurable infrared hyperbolic metasurfaces using phase change materials. *Nature communications* **9**, 1-7 (2018).
20      Chaudhary, K. *et al.* Polariton nanophotonics using phase-change materials. *Nature communications* **10**, 1-6 (2019).
21      Giles, A. J. *et al.* Imaging of anomalous internal reflections of hyperbolic phonon-polaritons in hexagonal boron nitride. *Nano letters* **16**, 3858-3865 (2016).
22      Alfaro-Mozaz, F. *et al.* Nanoimaging of resonating hyperbolic polaritons in linear boron nitride antennas. *Nature communications* **8**, 1-8 (2017).

23	Li, P. *et al.* Infrared hyperbolic metasurface based on nanostructured van der Waals materials. *Science* **359**, 892-896 (2018).
24	Giles, A. J. *et al.* Ultralow-loss polaritons in isotopically pure boron nitride. *Nature materials* **17**, 134 (2018).
25	Vuong, T. *et al.* Isotope engineering of van der Waals interactions in hexagonal boron nitride. *Nature materials* **17**, 152 (2018).
26	Kretinin, A. *et al.* Electronic properties of graphene encapsulated with different two-dimensional atomic crystals. *Nano letters* **14**, 3270-3276 (2014).
27	Liu, S. *et al.* Single crystal growth of millimeter-sized monoisotopic hexagonal boron nitride. *Chemistry of Materials* **30**, 6222-6225 (2018).
28	Fei, Z. *et al.* Gate-tuning of graphene plasmons revealed by infrared nano-imaging. *Nature* **487**, 82-85 (2012).
29	Chen, J. *et al.* Optical nano-imaging of gate-tunable graphene plasmons. *Nature* **487**, 77-81 (2012).
30	Dai, S. *et al.* Efficiency of launching highly confined polaritons by infrared light incident on a hyperbolic material. *Nano letters* **17**, 5285-5290 (2017).
31	Snyder, A. W. & Love, J. *Optical waveguide theory*.  (Springer Science & Business Media, 2012).
32	Caldwell, J. D. *et al.* Sub-diffractional, Volume-confined Polaritons in the Natural Hyperbolic Material Hexagonal Boron Nitride. *Nature Communications* **5**, 5221 (2014).
33	Bogaerts, W. *et al.* Silicon microring resonators. *Laser Photonics Reviews* **6**, 47-73 (2012).
34	Yoo, D. *et al.* High-contrast infrared absorption spectroscopy via mass-produced coaxial zero-mode resonators with sub-10 nm gaps. *Nano letters* **18**, 1930-1936 (2018).
35	Autore, M. *et al.* Boron nitride nanoresonators for phonon-enhanced molecular vibrational spectroscopy at the strong coupling limit. *Light: Science &Amp; Applications* **7**, 17172(2018).
36	Folland, T. G. *et al.* Vibrational Coupling to Epsilon-Near-Zero Waveguide Modes. *Acs Photonics* **7**, 614-621 (2020).
37	Berte, R. *et al.* Sub-nanometer Thin Oxide Film Sensing with Localized Surface Phonon Polaritons. *Acs Photonics* **5**, 2807-2815 (2018).
38	Neuner III, B. *et al.* Midinfrared index sensing of pL-scale analytes based on surface phonon polaritons in silicon carbide. *The Journal of Physical Chemistry C* **114**, 7489-7491 (2010).
39	Fali, A. *et al.* Refractive Index-Based Control of Hyperbolic Phonon-Polariton Propagation. *Nano letters* **19**, 7725-7734 (2019).